\documentclass[a4paper,11pt]{article}
\usepackage{pos}
\usepackage{bm}
\usepackage[export]{adjustbox}
\usepackage{graphicx}
\usepackage{xcolor}
\usepackage{xparse,xcoffins}

\ExplSyntaxOn
\NewCoffin\imagecoffin
\NewCoffin\labelcoffin

\keys_define:nn { miguel/label }
 {
  label   .tl_set:N = \l_miguel_label_tl,
  labelbox .bool_set:N = \l_miguel_label_box_bool,
  labelbox .default:n = true,
  fontsize .tl_set:N = \l_miguel_label_size_tl,
  fontsize .initial:n = \footnotesize,
  pos .choice:,
  pos/nw .code:n = \tl_set:Nn \l_miguel_label_pos_tl { left,up },
  pos/ne .code:n = \tl_set:Nn \l_miguel_label_pos_tl { right,up },
  pos/sw .code:n = \tl_set:Nn \l_miguel_label_pos_tl { left,down },
  pos/se .code:n = \tl_set:Nn \l_miguel_label_pos_tl { right,down },
  pos/n .code:n = \tl_set:Nn \l_miguel_label_pos_tl { hc,up },
  pos/w .code:n = \tl_set:Nn \l_miguel_label_pos_tl { left,vc },
  pos/s .code:n = \tl_set:Nn \l_miguel_label_pos_tl { hc,down },
  pos/e .code:n = \tl_set:Nn \l_miguel_label_pos_tl { right,vc },
  pos .initial:n = nw,
  unknown .code:n   = \clist_put_right:Nx \l_miguel_label_clist
                       { \l_keys_key_tl = \exp_not:n { #1 } }
 }
\clist_new:N \l_miguel_label_clist
\box_new:N \l_miguel_label_box
\box_new:N \l_miguel_label_image_box

\NewDocumentCommand{\xincludegraphics}{O{}m}
 {
  \group_begin:
  \tl_clear:N \l_miguel_label_tl
  \clist_clear:N \l_miguel_label_clist
  \keys_set:nn { miguel/label } { #1 }
  \tl_if_empty:NTF \l_miguel_label_tl
   {
    \miguel_includegraphics:Vn \l_miguel_label_clist { #2 }
   }
   {
    \SetHorizontalCoffin\imagecoffin
     {
      \miguel_includegraphics:Vn \l_miguel_label_clist { #2 }
     }
    \SetHorizontalCoffin\labelcoffin
     {
      \raisebox{\depth}
       {
        \bool_if:NTF \l_miguel_label_box_bool
         { \fcolorbox{white}{white}{\l_miguel_label_size_tl\l_miguel_label_tl} }
         { \l_miguel_label_size_tl\l_miguel_label_tl }
       }
     }
    \SetVerticalPole\imagecoffin{left}{0pt+\CoffinWidth\labelcoffin/2}
    \SetVerticalPole\imagecoffin{right}{\Width-3pt-\CoffinWidth\labelcoffin/2}
    \SetHorizontalPole\imagecoffin{up}{\Height-3pt-\CoffinHeight\labelcoffin/2}
    \SetHorizontalPole\imagecoffin{down}{3pt+\CoffinHeight\labelcoffin/2}
    \use:x{\JoinCoffins\imagecoffin[\l_miguel_label_pos_tl]\labelcoffin[vc,hc]} 
    \TypesetCoffin\imagecoffin
   }
   \group_end:
 }
\NewDocumentCommand{\setlabel}{m}
 {
  \keys_set:nn { miguel/label } { #1 }
 }

\cs_new_protected:Nn \miguel_includegraphics:nn
 {
  \includegraphics[#1]{#2}
 }
\cs_generate_variant:Nn \miguel_includegraphics:nn { V }

\ExplSyntaxOff

\def\1s0{^1 \hskip -0.03in S_0}
\def\3s1{^3 \hskip -0.025in S_1}

\title{Towards robust constraints on nuclear effective field theory from lattice QCD}

\manuallySeparateAuthors
\author*{Marc Illa}
\affiliation{Departament de Física Quàntica i Astrofísica, Institut de Ciències del Cosmos (ICCUB), Facultat de Física, Universitat de Barcelona, Martí i Franqués, 1, E08028 Barcelona, Spain}
\affiliation{InQubator for Quantum Simulation (IQuS), Department of Physics, University of Washington, Seattle, WA 98195, USA}

\author[1]{ for the NPLQCD collaboration}
\note{I want to acknowledge the rest of the members of the collaboration that have contributed to the work presented here: Silas R. Beane, Emmanuel Chang, Zohreh Davoudi, William Detmold, David J. Murphy, Patrick Oare, Kostas Orginos, Assumpta Parreño, Martin J. Savage, Phiala E. Shanahan, Brian C. Tiburzi, Michael L. Wagman, and Frank Winter.}

\emailAdd{marcilla@uw.edu}

\abstract{We will discuss several new results from the NPLQCD Collaboration that combine lattice QCD results on (hyper)nuclear systems at unphysical pion masses together with nuclear effective field theories. Two-baryon channels with strangeness $0$ to $-4$ are analyzed, with findings that point to interesting symmetries observed in hypernuclear forces as predicted in the limit of QCD with a large number of colors. Also, several matrix elements of light nuclei are studied. The tritium axial charge, related to the Gamow-Teller matrix element, and the longitudinal momentum fraction of $^3$He that is carried by the isovector combination of $u$ and $d$ are extracted and extrapolated to the physical point. For this latter case, it can be seen how including lattice results to experimental determinations can have imminent potential to enable more precise determinations and to reveal the QCD origins of the EMC effect.}

\FullConference{%
 The 38th International Symposium on Lattice Field Theory, LATTICE2021
  26th-30th July, 2021
  Zoom/Gather@Massachusetts Institute of Technology
}

\begin{document}
\maketitle

\section{Introduction}

The understanding of nuclear physics directly from the fundamental theory, Quantum Chromodynamics (QCD), is a difficult task due to the nonperturbative nature of QCD in the low-energy regime.
Several techniques are available, like the use of effective field theories (EFT)~\cite{Furnstahl:2021rfk}, that require data to constrain the low-energy coefficients (LECs), historically coming from experiments, or more recently from lattice QCD.

There are several advantages to using lattice QCD. In some instances, it is very difficult or impossible to extract the information experimentally. In others, it can be used to test the Standard Model (SM) and find deviations that might signal possible extensions.
In the following sections, we will discuss several new works related to these topics: the study of two octet baryons~\cite{NPLQCD:2020lxg}, the calculation of the axial charge of the triton~\cite{Parreno:2021ovq}, and the determination of the fraction of the longitudinal momentum of $^3$He that is carried by the isovector combination of $u$ and $d$ quarks~\cite{Detmold:2020snb}.

\section{Baryon-baryon interactions}

A profound knowledge of the interaction between baryons, especially when strangeness is involved, is necessary to achieve a correct description of, for example, the interior of neutron stars. However, since hyperons are unstable against the weak interaction, their experimental study is very challenging. 
In a lattice QCD calculation one can circumvent experimental limitations by switching off the weak interaction and retaining only the strong force, allowing the hyperons to be stable.


The present study follows the extensive work performed by the NPLQCD Collaboration at the $SU(3)$ flavor symmetric point~\cite{Wagman:2017tmp,NPLQCD:2012mex}. Here we extend the study at $m_\pi\sim 450$ MeV~\cite{Orginos:2015aya}, where only the two-nucleon systems were studied, by including nine different channels~\cite{NPLQCD:2020lxg}.
Using the notation $(^{2s+1} L_J,\, I)$, where $s$ is the total spin, $L$ is the orbital momentum, $J$ is the total angular momentum, and $I$ is the isospin, they are
\begin{align*}
 S=\phantom{-}0\; &: \; NN \;(\1s0,\, I=1), \ NN \;(\3s1,\, I=0), \\
 S=-1\; &: \; \Sigma N \;(\1s0,\, I=\tfrac{3}{2}),\ \Sigma N (\3s1,\, I=\tfrac{3}{2}),\\
 S=-2\; &: \; \Sigma \Sigma \;(\1s0,\, I=2),\ \Xi N \;(\3s1,\, I=0), \\
 S=-3\; &: \; \Xi \Sigma \;(\1s0,\, I=\tfrac{3}{2}),\\
 S=-4\; &: \; \Xi\Xi \;(\1s0,\, I=1), \ \Xi\Xi \;(\3s1,\, I=0).
\end{align*}
We briefly comment here on the constraints on the relevant LECs that appear in the EFT Lagrangians. We refer to Ref.~\cite{NPLQCD:2020lxg} for details on the extraction of the energy levels, phase-shifts and scattering parameters. 
\begin{itemize}
    \item Assuming $SU(3)$ flavor symmetry~\cite{Savage:1995kv,Petschauer:2013uua}, there are 6 LECs at leading order (LO) ($c_1,\ldots,c_6$) and 6 LECs at next-to-leading order (NLO) ($\tilde{c}_1,\ldots,\tilde{c}_6$) with terms that preserve $SU(3)$ flavor symmetry, and therefore, they can be grouped into a LEC for each irreducible representation of the product of two octets ($\mathbf{27}, \mathbf{8}_s, \mathbf{1},\overline{\mathbf{10}}, \mathbf{10}, \mathbf{8}_a$). Additionally, at NLO, there are 12 LECs with terms that explicitly break $SU(3)$ flavor symmetry ($c_1^\chi,\ldots,c_{12}^\chi$).
    \item Assuming $SU(6)$ spin-flavor symmetry~\cite{Kaplan:1998tg}, there are only 2 LECs, $a$ and $b$, at LO
\end{itemize}
These coefficients can be extracted by matching them to a momentum expansion of the scattering amplitude~\cite{Kaplan:1998tg,Kaplan:1998we,vanKolck:1998bw},
\begin{equation}
    \left[-\frac{1}{a_{\scriptscriptstyle B_1B_2}}+\mu\right]^{-1}= \frac{\overline{M}_{\scriptscriptstyle B_1B_2}}{2\pi}(c^{(\text{irrep})}+\bm{c}^{\chi}_{\scriptscriptstyle B_1B_2})\, ,
\end{equation}
where $c^{(\text{irrep})}$ stands for the appropriate linear combinations of the $c_i$ LECs (similarly for $\bm{c}^{\chi}_{\scriptscriptstyle B_1B_2}\equiv c^{\chi}_{\scriptscriptstyle B_1B_2}(m^2_K-m^2_{\pi})$, where $c^{\chi}_{\scriptscriptstyle B_1B_2}$ are linear combinations of the $c_i^\chi$ LECs), the variable $a_{\scriptscriptstyle B_1B_2}$ is the scattering length of the channel $B_1 B_2$, and $\overline{M}_{\scriptscriptstyle B_1B_2}$ is the reduced mass of that system.
The renormalization scale $\mu$ depends on the naturalness of the interactions, and here we will only show the results for the unnatural case, setting $\mu=m_{\pi}$.
Two sets of inputs can be used to constrain the numerical values for the LECs: 1) the scattering parameters $\{a^{-1},r\}$ obtained from two-parameter effective range expansion fits (method~I), and 2) the binding momenta can be used to compute the corresponding scattering length, related at LO by \mbox{$-a^{-1}+\kappa^{(\infty)}=0$} (method~II). 
\begin{figure}[tb]
\centering
\xincludegraphics[width=0.48\textwidth,label=(i)]{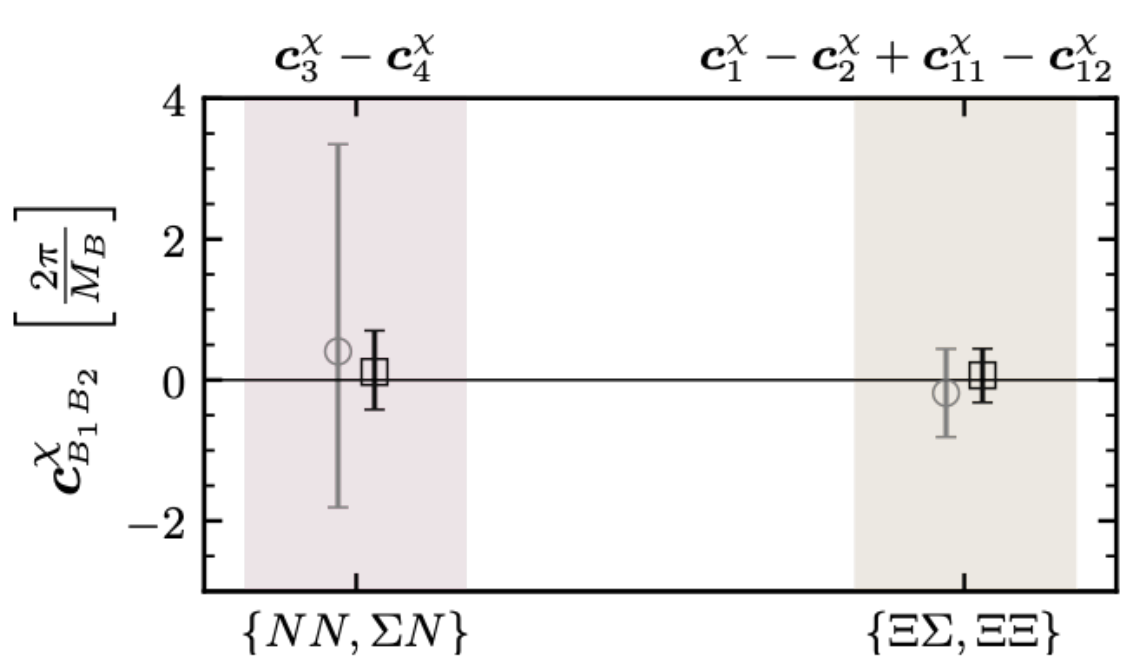}\hfill
\xincludegraphics[width=0.48\textwidth,label=(ii)]{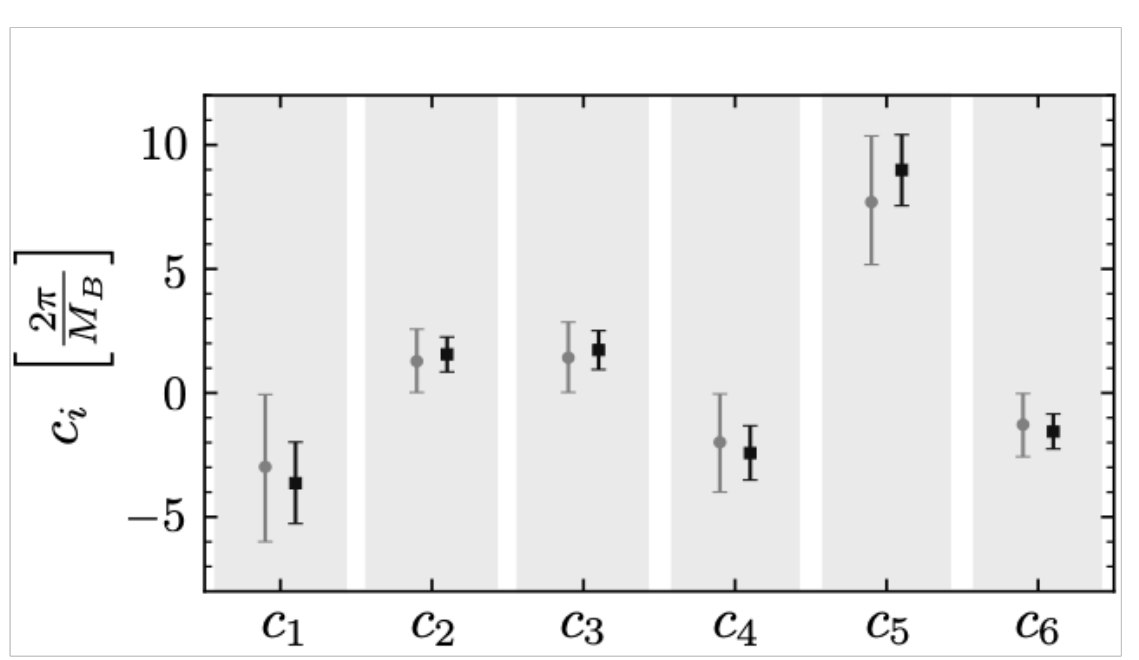}
\caption{(i) The NLO LECs $\bm{c}^{\chi}_{\scriptscriptstyle B_1B_2}$, in units of $[\frac{2\pi}{M_{B}}]$, where $M_B$ is the centroid of the octet-baryon masses, and (ii) the LO LECs $c_i$ reconstructed from the $SU(6)$ coefficients. The gray-circle markers denote quantities that are extracted using method I, while black-square markers show results obtained from method II.}
\label{fig:BB_LECs}
\end{figure}

As it can be seen from the left panel (i) of Fig.~\ref{fig:BB_LECs}, the symmetry breaking coefficients accessible in this work are compatible with zero, signaling that $SU(3)$ flavor symmetry remains an approximate symmetry, although we have set the light quarks to have a different mass than the strange quark. 
With this result, we are able to constrain the $SU(6)$ $a$ and $b$ LECs, and with those, compute all the $SU(3)$ LO LECs, shown in the right panel (ii) of Fig.~\ref{fig:BB_LECs}.
The relative importance of $c_5$ is a remnant of an accidental approximate $SU(16)$ symmetry of $s$-wave two-baryon interactions that is more pronounced in the $SU(3)$-symmetric study with $m_{\pi}\sim 806$ MeV in Ref.~\cite{Wagman:2017tmp}.

\section{Triton axial charge}

The triton isovector axial charge is very well known experimentally, with its ratio to the proton axial charge being $g^{^3\text{H}}_A/g^p_A=0.9511(13)$~\cite{Baroni:2016xll}.
This deviation from unity, more pronounced in heavier nuclei, has been historically elusive to be described directly from the SM. 
In our recent work~\cite{Parreno:2021ovq}, we use several ensembles with a pion mass of $m_\pi\sim 450$ MeV, and with a previous determination at $m_{\pi}\sim 806$ MeV~\cite{Savage:2016kon,Chang:2017eiq}, an extrapolation to the physical point is performed and compared to the physical value.

\begin{figure}[tb]
\centering
\xincludegraphics[width=0.44\textwidth,label=(i),valign=t]{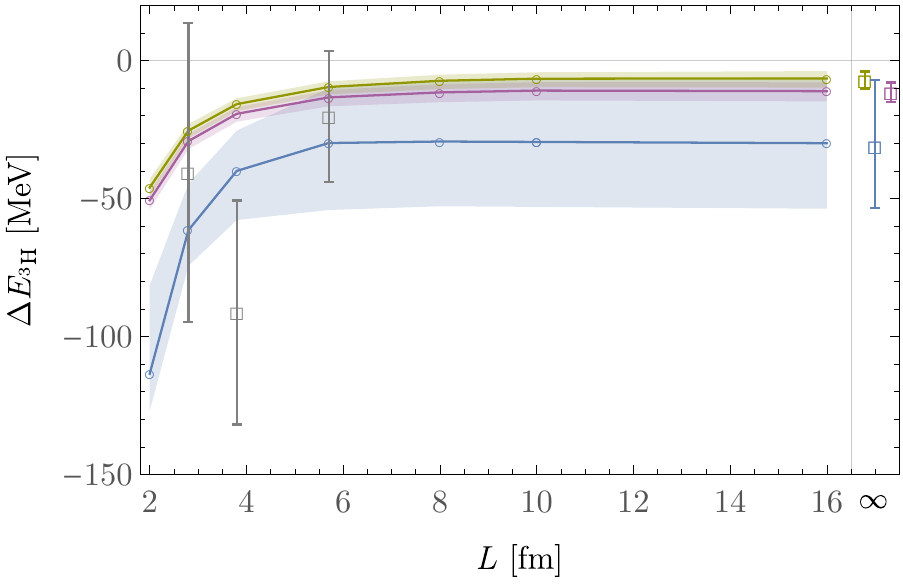}\hfill
\xincludegraphics[width=0.44\textwidth,label=(ii),valign=t]{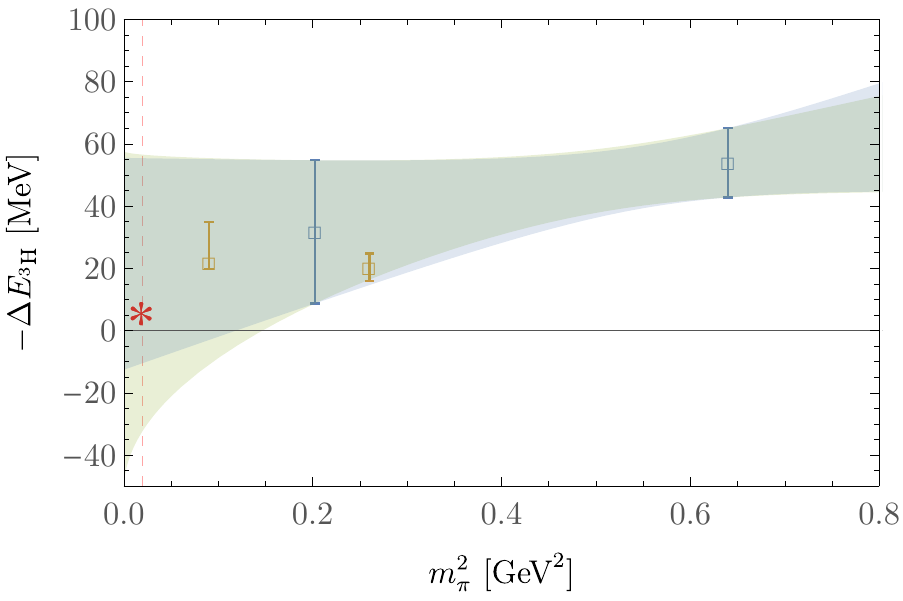}
\xincludegraphics[width=0.44\textwidth,label=(iii),valign=t]{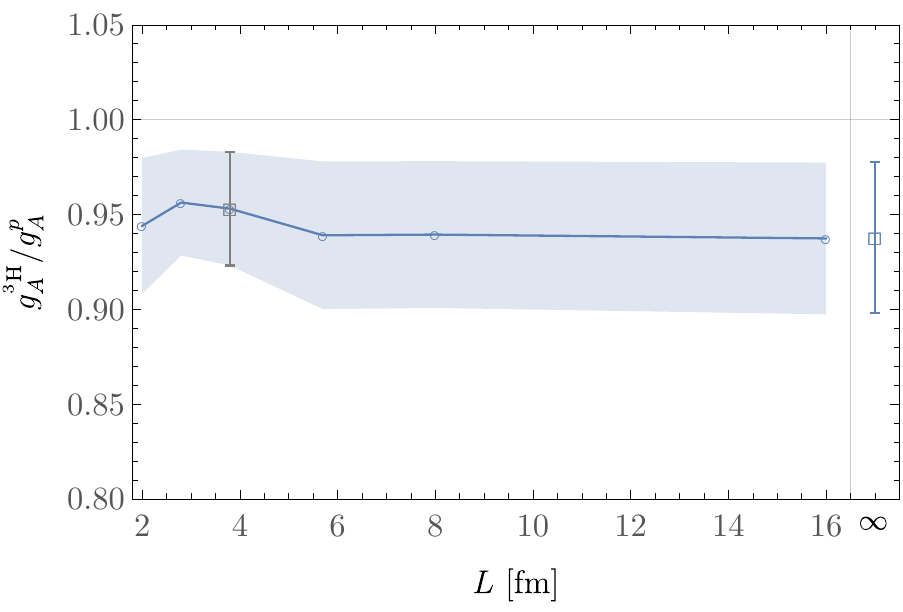}\hfill
\xincludegraphics[width=0.44\textwidth,label=(iv),valign=t]{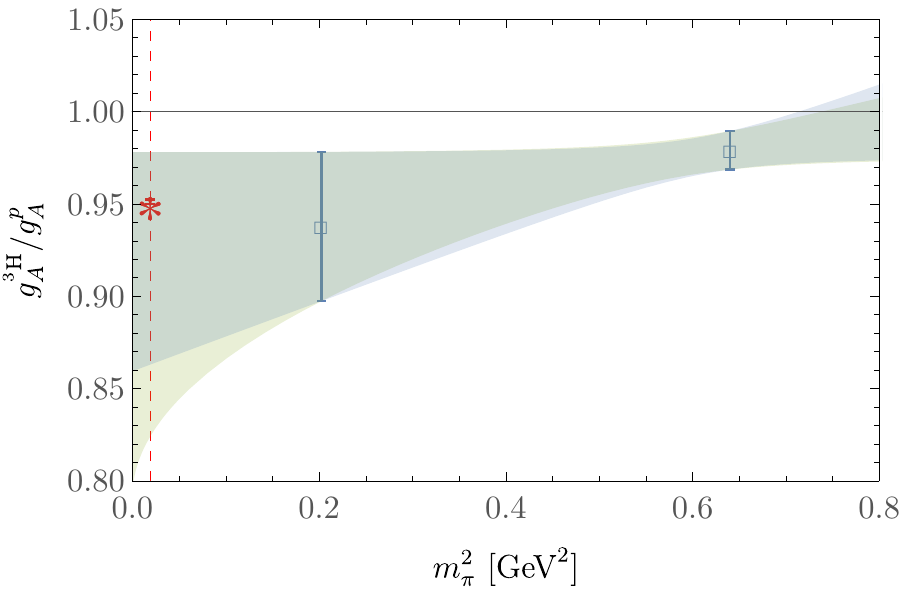}
\caption{(i) Finite-volume extrapolation of the ground-state energy of the triton (blue band) together with the deuteron (green) and dineutron (purple). (ii) Pion-mass extrapolation of the triton binding energy. (iii) Finite-volume extrapolation of the ratio of triton to proton axial charges. (iv) Pion-mass extrapolation of the ratio of triton to proton axial charges.}
\label{fig:3H_Axial}
\end{figure}

In order to extract the isovector axial charge, first the ground-state energy of the triton has to be computed.
Three different volumes were used, and in order to extrapolate the binding energy to the infinite-volume limit, since analytical forms are still not available, a numerical approach was taken using finite-volume EFTs (FVEFT)~\cite{Eliyahu:2019nkz}, where the LECs of the corresponding EFT are fixed used the lattice finite-volume results.
As shown in the top panels of Fig.~\ref{fig:3H_Axial}, the left one (i) shows volume dependence of the ground-state energy, with the blue band obtained with the FVEFT formalism, together with the lattice points (gray), for the triton, as well as the bands for the deuteron (green) and dineutron (purple).
Due to the large uncertainties in the three-body binding energy, the possibility that it is a $2+1$-body system rather than a compact 3-body system is not ruled out, although unlikely, and for the rest of the discussion we assume the latter.
The right panel (ii) shows the extrapolation of the binding energy to the physical pion mass using the current point at $m_\pi\sim 450$ MeV together with the one with heavier quark masses, at $m_\pi\sim 806$ MeV~\cite{NPLQCD:2012mex}, with two functional forms: linear (green) and quadratic (blue) in $m_\pi$
The physical point is shown as the red star, together with lattice calculations at similar pion masses for comparison~\cite{Yamazaki:2012hi,Yamazaki:2015asa}.

The next step is computing the matrix element $\langle ^3\text{H} | \overline{q}\gamma_3\gamma_5 \tau_3 q| ^3\text{H} \rangle$, with $\tau_a$ being a Pauli matrix in flavor space, using the background-field approach for building the required 3-point functions~\cite{Savage:2016kon}.
Although only one volume was used in this calculation, FVEFT can still be used to extrapolate to infinite volume~\cite{Detmold:2021oro}, as shown in the bottom left panel (iii) of Fig.~\ref{fig:3H_Axial}. Due to the large binding energy of the system, very mild volume dependence is observed, as expected. 
Together with the ratio computed at $m_\pi\sim 806$ MeV~\cite{Savage:2016kon,Chang:2017eiq}, an extrapolation is performed (with also linear and quadratic forms in $m_\pi$), and the resulting value, $g^{^3\text{H}}_A/g^p_A=0.91^{+0.07}_{-0.09}$, is found to be consistent with the experimental one, albeit with large uncertainties.

\section{Momentum fraction of \texorpdfstring{$^3$}{3}He}

Another quantity in the few-body sector that is of interest to determine directly from QCD is the modification of the parton distribution functions (PDFs) of nuclei compared to that of the single nucleons, known as the EMC effect~\cite{EuropeanMuon:1983wih}.
Although it is possible to determine the PDFs $q(x)$ directly from QCD, as it has been done in the single-hadron sector, in Ref.~\cite{Detmold:2020snb} we compute the first moment of the isovector quark PDFs, also known as momentum fraction, $\langle x \rangle_{u-d}=\int_{-1}^1 dx \, x \, [u(x)-d(x)]$, of $^3$He for the first time at $m_\pi\sim 806$ MeV, where the background field technique~\cite{Savage:2016kon} can also be applied.

\begin{figure}[tb]
\centering
\xincludegraphics[width=0.36\textwidth,label=(i),valign=t]{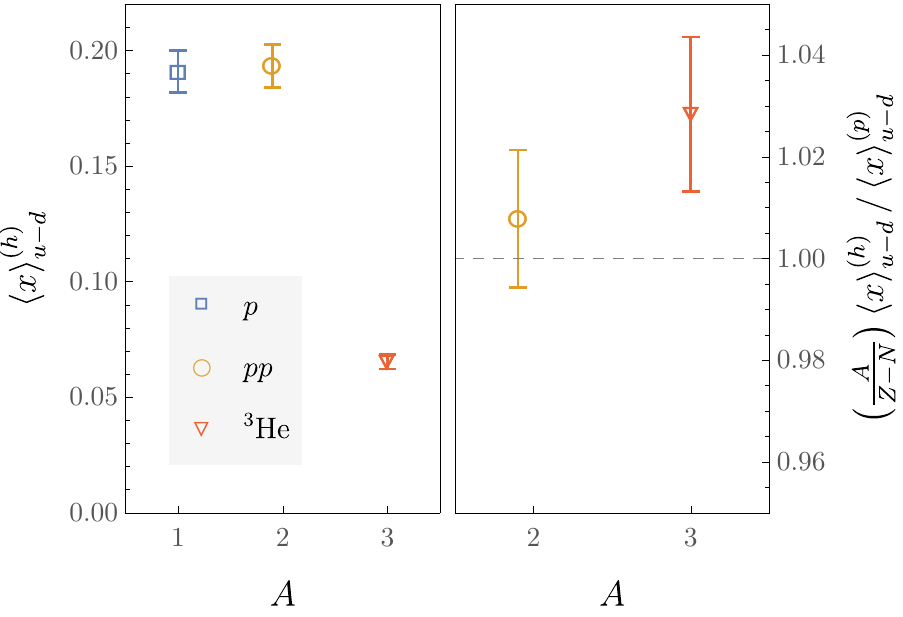}\hfill
\xincludegraphics[width=0.24\textwidth,label=(ii),valign=t]{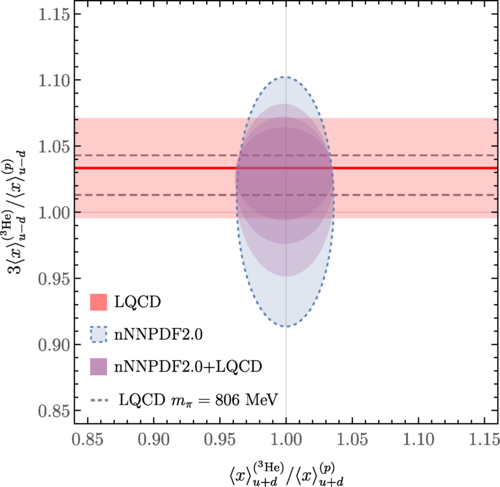}\hfill
\xincludegraphics[width=0.36\textwidth,label=(iii),valign=t]{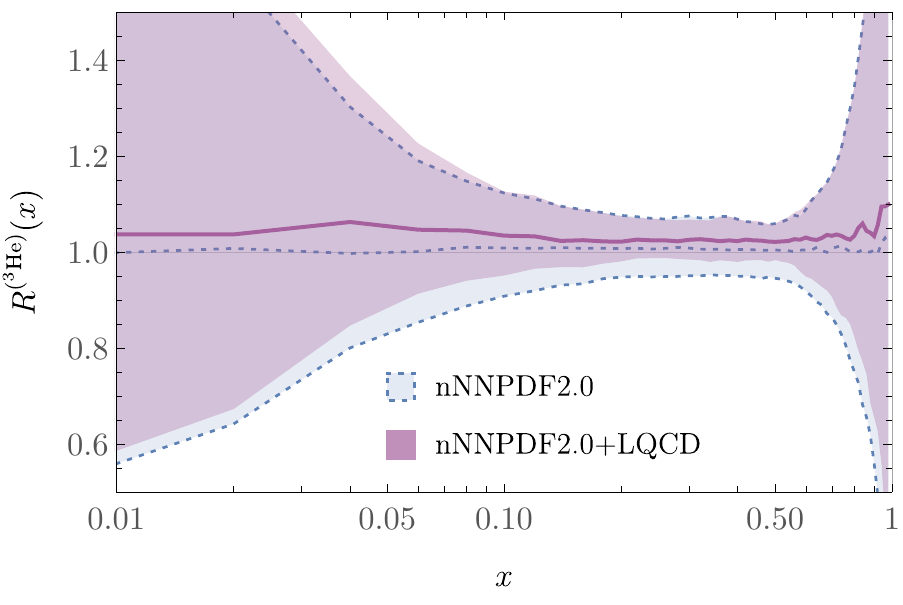}
\caption{(i) On the left, renormalized values of the isovector momentum fraction for $h\in\{p,pp,{}^3\text{He}\}$ at a scale of $\mu=2$~GeV, and on the right, ratios of the isovector nuclear momentum fractions to that of the constituent nucleons. (ii) The ratio of the isovector momentum fractions of ${}^3{\rm He}$ and $p$ determined in this work compared to constraints on the isovector and isoscalar momentum fraction ratios from the nNNPDF2.0~\cite{AbdulKhalek:2020yuc}  global analysis. (iii) The ratio of the isovector PDFs of ${}^3{\rm He}$ and $p$.}
\label{fig:Momfrac}
\end{figure}

The operator that we study is the twist-two operator~\cite{Gockeler:1996mu} $\mathcal{T}=(\mathcal{T}_{33}-\mathcal{T}_{44})/\sqrt{2}$, where $\mathcal{T}_{\mu\nu}=\overline{q}\tau_3\gamma_{\{\mu}\overleftrightarrow{D}_{\nu\}}q$, with $D_\mu$ being the gauge covariant derivative and $\{\ldots\}$ indicating symmetrization and trace subtraction of the enclosed indices. The results are shown in the first panel (i) of Fig.~\ref{fig:Momfrac} for all the systems studied: proton, diproton and $^3$He (for the nonperturbative renormalization of this operator, see the Supplemental Material of Ref.~\cite{Detmold:2020snb}).

Like in the case of the axial charge, the quantities in Ref.~\cite{Detmold:2020snb} have been computed using only one single volume, but with FVEFT~\cite{Detmold:2021oro}, they can be extrapolated to infinite volume. Now, however, there is no additional lattice calculation to help perform the extrapolation to the physical quark masses. In order to give an estimate, the following assumption is adopted: the product of the LEC $\alpha_{3,2}$ and nuclear factor $\mathcal{G}_3({}^3\text{He})$ is expected to have a mild quark-mass dependence. This product can be related to the momentum fraction as~\cite{Chen:2004zx}
\begin{equation}
    \alpha_{3,2}\mathcal{G}_3({}^3\text{He})=\frac{1}{3}\left(3\frac{\langle x \rangle_{u-d}^{({}^3\text{He})}}{\langle x \rangle_{u-d}^{(p)}}-1\right)\langle x \rangle_{u-d}^{(p)}\, .
\end{equation}
So, if $\alpha_{3,2}\mathcal{G}_3({}^3\text{He})$ is first fixed at $m_\pi\sim 806$ MeV with the lattice results, then, the same value, together with the phenomenological value of $\langle x \rangle_{u-d}^{(p)}$, can be used to extract the ratio $3 \langle x \rangle_{u-d}^{({}^3\text{He})}/\langle x \rangle_{u-d}^{(p)}$ at the physical point. This is shown in the panel (ii) of Fig.~\ref{fig:Momfrac}, where the dashed lines represent the lattice result at $m_\pi\sim 806$ MeV, and the red band the extrapolated value. In order to account for the unknown quark-mass dependence, an additional systematic uncertainty of the 100\% value of the LEC itself is added in quadrature, resulting in a value of $3 \langle x \rangle_{u-d}^{({}^3\text{He})}/\langle x \rangle_{u-d}^{(p)}=1.033(38)$.

Comparing this value with the one extracted from global nuclear PDF fits using the nNNPDF2.0 framework~\cite{AbdulKhalek:2020yuc}, shown as the blue ellipse in panel (ii) of Fig.~\ref{fig:Momfrac}, it seems we can use our lattice determination to further constrain the ratio via Bayesian reweighting~\cite{Ball:2011gg}, shown as the purple ellipses (the different sizes correspond to different \% values of the additional systematic uncertainty, \{50,100,200\}\%). With the same reweighting procedure, we can also reduce the uncertainty on the $x$-dependent PDFs themselves, as shown in panel (iii) of Fig.~\ref{fig:Momfrac}. Therefore, even with the most conservative estimate of the uncertainty, this shows how lattice QCD calculations can help improve our knowledge of nuclear PDFs.

\section{Summary}

We have shown here that lattice QCD can be used to extract information of nuclear systems, and make an impact in some long-standing problems. Further work is required, reaching pion mass values closer to the physical point, using several lattice spacings to understand discretization effects, as well as understanding the discrepancies of the energies extracted with different operators~\cite{Amarasinghe:2021lqa}, discussed in a talk given in this conference by M.~L.~Wagman~\cite{Wagman:2021}.

\section*{Acknowledgments}

The results presented in this manuscript were obtained using ensembles of isotropic-clover gauge-field configurations produced several years ago with resources obtained by researchers at the College of William and Mary and the Thomas Jefferson National Accelerator Facility and by the NPLQCD collaboration. Computations were performed using a College of William and Mary led XSEDE and NERSC allocation, and NPLQCD PRACE allocations on Curie and MareNostrum, on LLNL machines, on the HYAK computational infrastructure at the Univeristy of Washington, and through ALCC allocations.
Calculations of propagators and their contractions were performed using computational resources provided by the Extreme Science and Engineering Discovery Environment, which is supported by National Science Foundation Grant No.\ \mbox{OCI-1053575}. 
This research used resources of the Oak Ridge Leadership Computing Facility at the Oak Ridge National Laboratory, which is supported by the Office of Science of the U.S.\ Department of Energy under Contract number \mbox{DE-AC05-00OR22725}, as well as facilities of the USQCD Collaboration, which are funded by the Office of Science of the U.S.\ Department of Energy, and the resources of the National Energy Research Scientific Computing Center (NERSC), a U.S.\ Department of Energy Office of Science User Facility operated under Contract No.\ \mbox{DE-AC02-05CH11231}. The authors thankfully acknowledge the computer resources at MareNostrum and the technical support provided by Barcelona Supercomputing Center (\mbox{RES-FI-2019-2-0032} and \mbox{RES-FI-2019-3-0024}). Parts of the calculations used the Chroma~\cite{Edwards:2004sx}, QLua~\cite{qlua}, QUDA~\cite{Clark:2009wm,Babich:2011np,Clark:2016rdz}, QDP-JIT~\cite{6877336} and QPhiX~\cite{10.1007/978-3-319-46079-6_30} software libraries. 

The author has been supported in part by the Universitat de Barcelona through the scholarship APIF, by the Spanish Ministerio de Economía y Competitividad (MINECO) under the project \mbox{MDM-2014-0369} of ICCUB (Unidad de Excelencia “María de Maeztu”), by the European FEDER funds under the contract \mbox{FIS2017-87534-P} and by the EU STRONG-2020 project under the program \mbox{H2020-INFRAIA-2018-1}, grant agreement No.\ 824093, and in part by the U.S. Department of Energy, Office of Science, National Quantum Information Science Research Centers, Quantum Science Center, and by the U.S. Department of Energy, Office of Science, Office of Nuclear Physics, InQubator for Quantum Simulation (IQuS) through the Quantum Horizons: QIS Research and Innovation for Nuclear Science, under Award Number DOE (NP) Award
\mbox{DE-SC0020970}.

\bibliographystyle{JHEP}
\bibliography{bibliography}

\end{document}